\documentclass[apjl,numberedappendix]{emulateapj}
\usepackage{mathptmx}
\usepackage{txfonts}
\usepackage[T1]{fontenc}
\usepackage{ae,aecompl}
\usepackage{booktabs}
\usepackage[flushleft]{threeparttable}

\usepackage{graphicx}
\usepackage{wasysym}
\usepackage{hyperref}
\usepackage{extdash}
\usepackage{multirow}
\usepackage[nolist]{acronym}

\def\addVAR#1#2{\expandafter\gdef\csname my@data@\detokenize{#1}\endcsname{#2}}
\def\VAR#1{%
  \ifcsname my@data@\detokenize{#1}\endcsname
    \csname my@data@\detokenize{#1}\expandafter\endcsname
  \else
    \expandafter\ERROR
  \fi
}

\addVAR{search.grb.snr|round(1)}{4.6}
\addVAR{search.assoc.ligo.sig|round(1)}{3.2}
\addVAR{search.assoc.gbm.sig|round(1)}{4.2}
\addVAR{optical_counterpart.name}{AT2017gfo}
\addVAR{thegrb.classification.prob_short_pc_min}{99}
\addVAR{triangulation.gbm_improvement_ipn|round(1)}{1.8}
\addVAR{triangulation.gbm_improvement_own|round(1)}{2.8}
\addVAR{(coverage_pointing.jemx_pointing_fraction_f20*100)|int}{95}
\addVAR{(coverage_pointing.jemx_pointing_fraction_f2*100)|int}{29}
\addVAR{table.spi[2000,4000].mcrab|int}{2117}
\addVAR{table.spi[2000,4000].flux_ecs|latex_exp}{5.7$\times$10$^{-9}$}
\addVAR{table.spi[2000,4000].L|latex_exp}{1.1$\times$10$^{45}$}
\addVAR{table.spi[4000,8000].mcrab|int}{5812}
\addVAR{table.spi[4000,8000].flux_ecs|latex_exp}{1.3$\times$10$^{-8}$}
\addVAR{table.spi[4000,8000].L|latex_exp}{2.5$\times$10$^{45}$}

\newif\ifdumbeddown
\dumbeddowntrue

\newif\ifpaper
\paperfalse

\newif\ifcomments
\commentsfalse

\hypersetup{
pdftitle={GW170817 and INTEGRAL},
pdfsubject={GW170817 and INTEGRAL},
pdfauthor={Savchenko et al},
pdfkeywords={GW170817, INTEGRAL}
}

\newcommand{\ecse}[2] {${#1} \times 10^{{#2}}$\,erg\,cm$^{-2}$\,s$^{-1}$}

\usepackage{color}

\ifcomments
\newcommand{\change}[1] {\textbf{#1}}
\newcommand{\ch}[1] {\textbf{#1}}

\else
\newcommand{\change}[1] {#1}
\newcommand{\ch}[1] {#1}

\fi

\newcommand{\software}[1] {\textit{#1}}

\begin{document}

\shorttitle{INTEGRAL observations of GW170817}
\shortauthors{V. Savchenko et al.}

\title{INTEGRAL Detection of the First Prompt Gamma-Ray Signal
  Coincident with the Gravitational Wave Event GW170817}

\author{ V.~Savchenko$^{1}$, 
C.~Ferrigno$^{1}$, 
E.~Kuulkers$^{2}$, 
A.~Bazzano$^{3}$, 
E.~Bozzo$^{1}$, 
S.~Brandt$^{4}$, 
J.~Chenevez$^{4}$, 
T.~J.-L.~Courvoisier$^{1}$, 
R.~Diehl$^{5}$,
A.~Domingo$^{6}$,
L.~Hanlon$^{7}$, 
E.~Jourdain$^{8}$
A.~von~Kienlin$^{5}$, 
P.~Laurent$^{9,10}$, 
F.~Lebrun$^{9}$, 
A.~Lutovinov$^{11,12}$, 
A.~Martin-Carrillo$^{7}$, 
S.~Mereghetti$^{13}$,
L. Natalucci$^{3}$,
J.~Rodi$^{3}$,
J.-P.~Roques$^{8}$, 
R.~Sunyaev$^{11,14}$, 
 and P.~Ubertini$^{3}$
\\
  }
\affil{  $^{1}$ISDC, Department of Astronomy, University of Geneva, Chemin d'\'Ecogia, 16 CH-1290 Versoix, Switzerland \\
$^{2}$European Space Research and Technology Centre (ESA/ESTEC), Keplerlaan 1, 2201 AZ Noordwijk, The Netherlands \\
 $^{3}$INAF-Institute for Space Astrophysics and Planetology, Via Fosso del Cavaliere 100, I-00133-Rome, Italy \\
 $^{4}$DTU Space, National Space Institute Elektrovej - Building 327 DK-2800 Kongens Lyngby Denmark \\
 $^{5}$Max-Planck-Institut f\"{u}r Extraterrestrische Physik, Garching, Germany \\
 $^{6}$Centro de Astrobiolog\'ia (CAB-CSIC/INTA, ESAC Campus), Camino bajo del Castillo S/N, E-28692 Villanueva de la Ca\~nada, Madrid, Spain\\
 $^{7}$Space Science Group, School of Physics, University College Dublin, Belfield, Dublin 4, Ireland \\
 $^{8}$IRAP, Universit\'e de Toulouse, CNRS, UPS, CNES; 9 Av. Roche, F-31028 Toulouse, France \\
 $^{9}$APC, AstroParticule et Cosmologie, Universit\'e Paris Diderot, CNRS/IN2P3, CEA/Irfu, Observatoire de Paris Sorbonne Paris Cit\'e,\\
\hspace{0.05cm}10 rue Alice Domont et L\'eonie Duquet, F-75205 Paris Cedex 13, France. \\
 $^{10}$DSM/Irfu/Service d'Astrophysique, Bat. 709 Orme des Merisiers CEA Saclay, F-91191 Gif-sur-Yvette Cedex, France \\
 $^{11}$Space Research Institute of Russian Academy of Sciences, Profsoyuznaya 84/32, 117997 Moscow, Russia  \\
 $^{12}$Moscow Institute of Physics and Technology, Institutskiy per. 9, 141700 Dolgoprudny, Moscow Region, 141700, Russia\\
 $^{13}$INAF, IASF-Milano, via E.Bassini 15, I-20133 Milano, Italy \\
 $^{14}$Max Planck Institute for Astrophysics, Karl-Schwarzschild-Str. 1, Garching b. D-85741, Munchen, Germany \\
  }

\date{Accepted XXX. Received YYY; in original form ZZZ}

\label{firstpage}

\begin{abstract}

  We report the e INTernational Gamma-ray Astrophysics Laboratory
  (INTEGRAL) detection of the short gamma-ray burst GRB~170817A
  \change{(discovered by Fermi-GBM) with a signal-to-noise ratio of
    \VAR{search.grb.snr|round(1)}}, and, for the first time, its
  association with the gravitational waves (GWs) from binary neutron star (BNS)
  merging event GW170817 detected by the LIGO and Virgo
  observatories. \change{The significance of association between the
    gamma-ray burst observed by INTEGRAL and GW170817 is
    \VAR{search.assoc.ligo.sig|round(1)}~$\sigma$, while the association
    between the Fermi-GBM and INTEGRAL detections is
    \VAR{search.assoc.gbm.sig|round(1)}~$\sigma$.  GRB~170817A was
    detected by the SPI-ACS instrument about 2~s after the end
    of the gravitational wave event. We measure a fluence of}
  $(1.4 \pm 0.4 \pm 0.6) \times$10$^{-7}$~erg~cm$^{-2})$
  (75--2000~keV), where, respectively, the statistical error is given
  at the 1~$\sigma$ confidence level, and the systematic error
  corresponds to the uncertainty in the spectral model and instrument
  response.

  We also report on the pointed follow-up observations carried out by
  INTEGRAL, starting 19.5~h after the event, and lasting for 5.4
  days.  We provide a stringent upper limit on any electromagnetic
  signal in a very broad energy range, from 3~keV to 8~MeV,
  constraining the soft gamma-ray afterglow flux to
  $<7.1\times$10$^{-11}$~erg~cm$^{-2}$~s$^{-1}$ (80--300\,keV).

  Exploiting the unique capabilities of INTEGRAL, we constrained the
  gamma-ray line emission from radioactive decays that are expected
  to be the principal source of the energy behind a kilonova event
  following a BNS coalescence.  Finally, we put a
  stringent upper limit on any delayed bursting activity, for example
  from a newly formed magnetar.

\end{abstract}

\section{Introduction}

\begin{acronym}
\acrodef{BH}[BH]{black hole}
\acrodef{BAT}[BAT]{Burst Alert Telescope}
\acrodef{BNS}[BNS]{binary neutron star}
\acrodef{EM}[EM]{electromagnetic}
\acrodef{EOS}[EOS]{equation of state}
\acrodef{FAP}[FAP]{false alarm probability}
\acrodef{FAR}[FAR]{false alarm rate}
\acrodef{GBM}[GBM]{Gamma-ray Burst Monitor}
\acrodef{GRB}[GRB]{gamma-ray burst}
\acrodef{sGRB}[sGRB]{short gamma-ray burst}
\acrodef{GW}[GW]{gravitational-wave}
\acrodef{INTEGRAL}[INTEGRAL]{INTErnational Gamma-Ray Astrophysics Laboratory}
\acrodef{IPN}[IPN]{InterPlanetary Network}
\acrodef{LIGO}[LIGO]{Laser Interferometer Gravitational-Wave Observatory}
\acrodef{MCMC}[MCMC]{Markov Chain Monte Carlo}
\acrodef{NS}[NS]{neutron star}
\acrodef{sGRB}[sGRB]{short gamma-ray burst}
\acrodef{SNR}[S/N]{signal-to-noise ratio}
\acrodef{SME}[SME]{Standard Model Extension}
\acrodef{SPI-ACS}[SPI-ACS]{SPectrometer onboard INTEGRAL - Anti-Coincidence Shield}
\end{acronym}

It has long been conjectured that the subclass of gamma-ray bursts
(GRBs) with a duration below about 2~s, known as \acp{sGRB},
are the product of a \ac{BNS} merger and that gamma-rays are produced
in the collimated ejecta following the coalescence
\citep[e.g,][]{blinnikov1984,Nakar2007,Gehrels2012,Berger2014}.  So
far, there was only circumstantial evidence for this hypothesis, owing
to the lack of supernovae associated with sGRBs, their localization in
early-type galaxies and their distinct class of duration
\citep[e.g.,][]{Davanzo2015}. The advent of advanced \ac{GW}
detectors, which have been able to detect binary black hole mergers
\citep{GW150914,LVT151012,GW151226,GW170104,GW170814}, and have the
capability to detect a signal from nearby BNS mergers
\citep{Abbott2016} have sparked great expectations.  Different
electromagnetic signatures are expected to be associated with BNS
merger events, owing to expanding ejecta, the most obvious of which is
an sGRB in temporal coincidence with the GW signal and/or afterglow
emission at different wavelengths in the days and/or weeks after the
merger event \citep[e.g.,][]{Fernandez2016}.

On n 2017 August 17 a at 12:41:04.47 UTC (T$_{0,GW}$ hereafter), a
signal consistent with the merger of a \ac{BNS} was detected by the
LIGO-Hanford detector \citep{aLIGO_instrument} \change{as a
  single-detector trigger.  The subsequent alert was issued} in
response to a public real-time Fermi \ac{GBM} trigger on a sGRB at
12:41:06.48 UTC \citep{GCN21505,GCN21506,GCN21509}; the GRB signal was
immediately and independently confirmed by our team \citep{GCN21507}.

Analysis of the LIGO-Livingston data \citep{LVC_PRL_GW170817} revealed
that a trigger was not automatically issued \change{due to the
  proximity of an overflow instrumental transient}, which could be
safely removed offline.  The addition of Virgo
\citep{TheVirgo:2014hva} to the detector network allowed a precise
localization at 90\% confidence level in an area of about 31 square
degrees \citep{GCN21513}, which is consistent with the Fermi-GBM
localization of GRB~170817A \citep{GCN21520}.  The most accurate
localization so far has been derived by the LALInferrence pipeline
\citep{GCN21527}; this is the localization we use in this Letter.

A massive follow-up campaign of the LIGO-Virgo high-probability region
by optical robotic telescopes started immediately after the event and
on 2017 August 18 between 1:05 and 1:45 UT, three groups reported
independent detections of a transient optical source at about
10~arcsec from the center of the host S0 Galaxy NGC~4993; this source
was dubbed SSS17a \citep{GCN21529,GCN21529_paper} or DLT17ck
\citep{GCN21531,GCN21531_paper}; the transient source was confirmed by
\citet{GCN21530} \citep[see also][]{GCN21530_paper}.  The source was
identified as the most probable optical counterpart of the BNS merger
\citep{GCN21557,GCN21557_paper}.  After that, it was followed at all
wavelengths. \ch{The counterpart has been given the official IAU
  designation ``AT2017gfo'' \citep{capstone}.}

\ch{During early observations by the Swift satellite from 53.8 to 55.8 ks
  after the LVC trigger an ultraviolet (UV) transient with $u$
  magnitude 17 was detected; an X-ray upper limit was set at an order of magnitude below
  the typical luminosity of an sGRB afterglow, as determined from 
  the sample of
  Swift \ac{BAT} triggered objects.  It was suggested that the object may be a
  blue (i.e., lanthanide-free) kilonova \citep{GCN21550}. Infrared
  spectroscopy with X-shooter on the ESO Very Large Telescope UT 2
  covered the wavelength range 3000--25000 \AA\ and started roughly
  1.5 days after the GW event \citep{GCN21592}. As reported in
  \citet{Pian2017}, strong evidence was found for $r$-process
  nucleosynthesis as predicted by kilonova models
  \citep[e.g.,][]{Tanaka2017}. Gemini spectroscopic observations with the
  Flamingos-2 instrument taken 3.5 days after the GW event revealed a red
  featureless spectrum, again consistent with kilonova
  expectations \citep{GCN21682,Troja2017}. Optical spectra collected
  1.5~days after the GW event with the ESO New Technology Telescope at La Silla
  equipped with the EFOSC2 instrument in spectroscopic mode excluded
  a supernova as being the origin of the transient emission
  \citep{GCN21582}.
Thus,} the properties of the source are fully consistent with the
a kilonova scenario \citep[see][for a
review]{Metzger2017}. A kilonova is primarily powered by the
radioactive decay of elements synthesized in the outflow, 
which produce gamma-ray
lines.  These may also be directly detectable in the gamma-ray range
\citep{Hotokezaka2016}.

In this Letter, we describe in detail the detection of GRB~170817A by
the INTernational Gamma-ray Astrophysics Laboratory (INTEGRAL) and the
targeted follow-up observing campaign. We were able to search for any
possible hard X-ray / soft gamma-ray emission for about six days after
the prompt gamma-ray and GW signal. This allowed us to constrain both
continuum emission from GRB-like afterglow emission and line emissions
expected from kilonovae.

\section{INTEGRAL instrument summary}
\label{sec:instrument}

INTEGRAL \citep{integral} is an observatory with multiple instruments:
a gamma-ray spectrometer (20\,keV--8\,MeV, SPI, \citealt{spi}), an
imager (15\,keV--10\,MeV, IBIS, \citealt{ibis}), an X-ray monitor (3--35\,keV, JEM-X, \citealt{jemx}), and an optical monitor (V band, OMC,\citealt{omc}). 

The spectrometer SPI is surrounded by a thick Anti-Coincidence Shield
(SPI-ACS).  In addition to its main function of providing a veto
signal for charged particles irradiating the SPI instrument, the ACS is
also able to register all other impinging
particles and high-energy photons.  Thus, it can be used as a nearly
omnidirectional detector of transient events with an effective area
reaching 0.7~m$^2$ at energies above $\sim$75~keV and a time
resolution of 50~ms \citep{spiacs}. The characterization of its
response to a gamma-ray signal has been delivered with an extensive
simulation study, taking into account the complex opacity pattern of
materials, which are used for the INTEGRAL satellite structure and
other instrument detectors.  Similarly, we have computed and verified
the response of the other omnidirectional detectors on board INTEGRAL:
IBIS/ISGRI, IBIS/PICsIT and IBIS/Veto. For details on the INTEGRAL
capabilities of detecting transients from the whole sky, particularily as relevant to our search for electro-magnetic counterparts to GW signals, we refer to \citet{Savchenko2017a} and references therein.

\section{Observation of the Prompt Gamma-Ray Emission}

\begin{figure}
\centering
  \includegraphics[width=1.\linewidth]{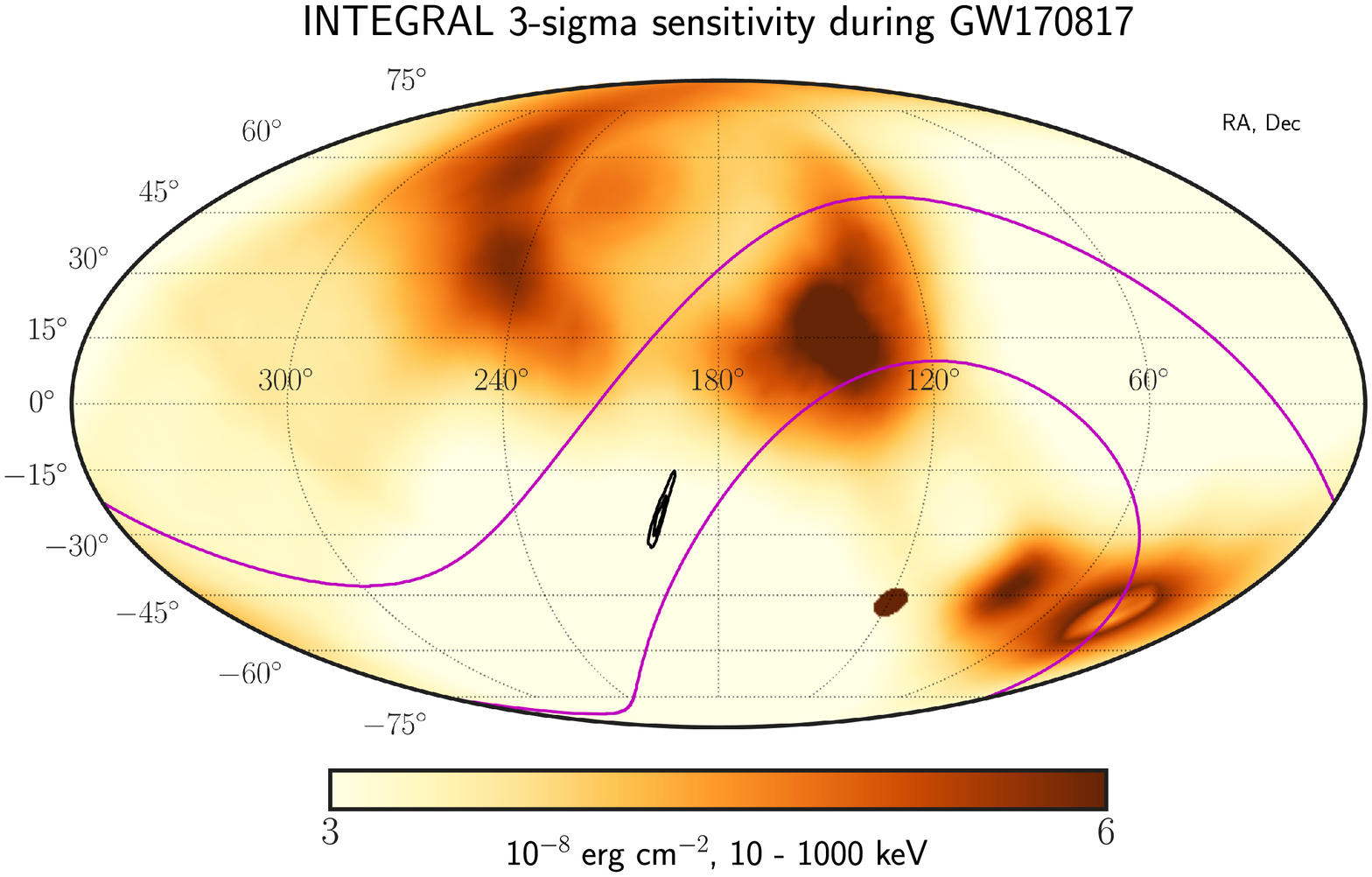}
  \caption{INTEGRAL 3~$\sigma$ sensitivity to a 100~ms burst characterized
    by Comptonized emission with $\alpha$=-0.65 and
    E$_{peak}$=185~keV, i.e., the best fit spectral model reported by
    the Fermi-GBM for the pulse of the GRB. Black contours correspond
    to the confidence regions (90\% and 50\%) of the current
    LALInferrence LIGO/Virgo localization \citep{GCN21527}. The magenta
    annulus corresponds to the constraint on the GRB~170817A location
    derived from the difference in arrival time of the event to Fermi
    and INTEGRAL \citep[triangulation;][]{GCN21515}}
\label{fig:skysens}
\end{figure}

At the time of the GW170817 trigger, INTEGRAL was performing a
target-of-opportunity observation of GW170814
\citep{GW170814,GCN21474} and at 12:41 UTC it was directed to RA, Dec
(J2000.0) = 36.25$^{\circ}$,\,-49.80$^{\circ}$.  This orientation was
overall favorable to detect a signal from the location of
\VAR{optical_counterpart.name} with the SPI-ACS, although not the most
optimal. This can be seen in Fig.~\ref{fig:skysens}, where we show
the complete INTEGRAL sensitivity map combining all instruments as
described in \citet{Savchenko2017a}.  We note that with this
orientation, the sensitivity of IBIS (ISGRI, \citealt{isgri}; PICsIT,
\citealt{picsit} and Veto, \citealt{veto} detectors) to a signal from
the direction of \VAR{optical_counterpart.name} was much lower if
compared to SPI-ACS for any plausible type of event spectrum.

We searched the SPI-ACS light curve using five timescales from 0.1 to
10~s, within a window of 30~s before and after the time of GW170817.
The local background noise properties are in good agreement with the
expectation for the background at the current epoch. On a 100~ms time
scale, we detect only one significant excess with a signal-to-noise
ratio (S/N) of \VAR{search.grb.snr|round(1)}, starting at
T$_{0,GW}$+1.9 s (in the geocentric time system; hereafter,
T$_{0,ACS}$). \ch{We compute a significance of association between
  GRB~170817A as observed by INTEGRAL and GW170817 of
  \VAR{search.assoc.ligo.sig|round(1)}~$\sigma$. The association
  significance with the Fermi-GBM observation of GRB~170817A is
  \VAR{search.assoc.gbm.sig|round(1)}~$\sigma$ (see
  Appendix~\ref{sec:association}).}

\begin{figure}
\centering
  \includegraphics[width=1.\linewidth]{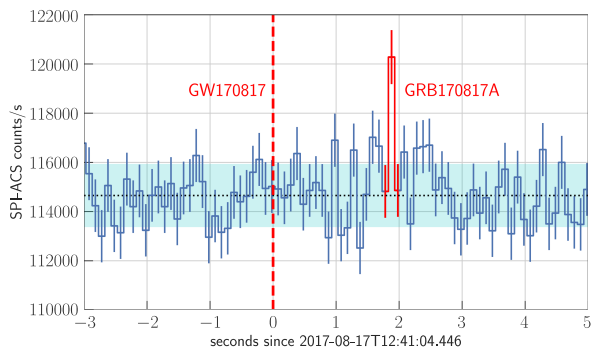}
  \caption{SPI-ACS light curve of GRB~170817A (100~ms time resolution),
    detected 2 seconds after GW170817. The red line highlights the
    100~ms pulse, \ch{which has an S/N of \VAR{search.grb.snr|round(1)}} in
    SPI-ACS. The blue shaded region corresponds to a range of one
    standard deviation of the background.}
\label{fig:acslc}
\end{figure}

\ch{The principal part of the excess gamma-ray emission emerges in
  just two 50\,ms time bins. The 100~ms duration firmly places this
  event in the short GRB class at
  $>$\VAR{thegrb.classification.prob_short_pc_min}\% probability
  \citep[using the GRB duration distribution
  from][]{Savchenko2012,Qin2013}. We should note, however, that the
  SPI-ACS does not have the capability to observe emission below
  $\sim$100~keV (due to the limitations of the instrumental low
  threshold), which might have slightly different temporal
  characteristics, as reported by \citet{gbm_only_paper}.}

\ch{Our coincident observation of the gamma-ray signal permits a
  substantial improvement of the Fermi-GBM-only localization by
  exploiting the difference in the gamma-ray arrival times at the
  location of the two satellites.  Using the triangulation annulus
  reported by \cite{GCN21515} we compute that the addition of the
  INTEGRAL observation reduces the final 90\% GBM localization area by
  a factor \VAR{triangulation.gbm_improvement_ipn|round(1)}.  We refer
  to the joint LIGO/Virgo, Fermi-GBM, and INTEGRAL/SPI-ACS analysis
  for more details
  \citep{joint_paper}. Appendix~\ref{sec:prompt_multiinstrument}
  summarizes the supporting complete INTEGRAL data set at the time of
  GRB~170817A.}

\begin{figure}
\centering
  \includegraphics[width=1.\linewidth]{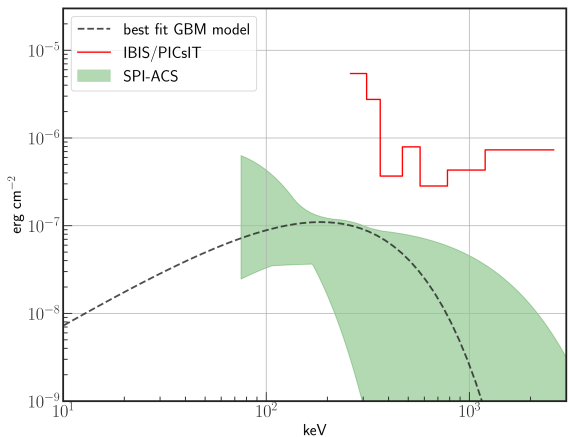}
  \caption{Average hard X-ray/gamma-ray spectrum of the initial pulse
    of GRB~170817A. The shaded green region corresponds to the range of
    spectra compatible with the INTEGRAL/SPI-ACS observation (see text
    for details).  IBIS/PICsIT provides a complementary independent
    upper limit at high energies; see text.  The best-fit Fermi-GBM
    model for the spectrum in the same interval (Comptonized model
    with low-energy index of -0.62 and $E_{peak}$ of 185\,keV) is
    shown as a dashed line for comparison \citep{gbm_only_paper}. }
\label{fig:prompt_spectra}
\end{figure}

The majority of \acp{sGRB} have a hard spectrum, resulting in a strong
detection in the SPI-ACS and/or in IBIS (ISGRI, PICsIT, and/or Veto),
as long as the respective instrument is favorably oriented
\citep{Savchenko2012, Savchenko2017a}. GRB~170817A, on the other hand,
was very soft, with most of its energy below $\sim$100\,keV, apart
from a short-hard initial pulse emitting at least up to 200~keV
\citep{gbm_only_paper}. This results in a reduced SPI-ACS signal
significance. We determined that for the location of
\VAR{optical_counterpart.name}, the SPI-ACS efficiency is smoothly
increasing from about 100 keV to 200 keV, where it reaches a plateau
up to the upper energy threshold of $\sim$80\,MeV. In
Figure~\ref{fig:prompt_spectra}, we show the region that contains the
allowed spectral models consistent with the SPI-ACS observation. We
assume a specific family of models, representative of sGRB spectra not
far from the Fermi-GBM best-fit model of GRB~170817A for time time
interval T$_{0,GBM}$-0.320~--~T$_{0,GBM}$+0.256--(covering the range
of spectra consistent with the average or hard peak):
Comptonized/cut-off power-law models with -1.7$\leq \alpha \leq$-0.2
and 50$\leq E_{peak} \leq$300~keV.  In the same figure, the black
dashed line represents the best-fit Fermi-GBM model in the same
0.576~s long time interval that we used to compare SPI-ACS with the
Fermi-GBM results \citep{gbm_only_paper}. This comparison nicely
displays the consistency of both experiments.

Due to the lack of energy resolution in SPI-ACS, the
fluence estimate depends on model assumptions. Using the best-fit
Fermi-GBM model \citep{gbm_only_paper} and assuming the time interval
T$_{GBM,0}$-0.320, T$_{GBM,0}$+0.256\,s to match the interval used by Fermi-GBM,
\citep{gbm_only_paper}, we derive a 75--2000 keV fluence of
$(1.4\pm 0.4)\times$10$^{-7}$~erg~cm$^{-2}$ (1~$\sigma$ error,
statistical only).  Additionally, the model assumption uncertainty
employing the same range of models as used in Figure~\ref{fig:prompt_spectra}
corresponds to a 75--2000~keV fluence uncertainty of
$\pm$0.4$\times$10$^{-7}$~erg~cm$^{-2}$. Possible systematic
deviations of the SPI-ACS response, as established by
cross-calibration with other gamma-ray instruments (primarily
Fermi-GBM and Konus-Wind), corresponds to a further uncertainty
of $\pm$0.3$\times$10$^{-7}$~erg~cm$^{-2}$.

Due to the limited duration of this event in SPI-ACS, little can be
learned directly from the light curve. However, we note that the main
prompt component consists of just two bins, with each of the rise time
and decay time below 50~ms. Our variability limits are derived for the
particularly narrow pulse that characterizes the hardest component of
the burst, which is observed by INTEGRAL/SPI-ACS with high effective
area. Our results should be compared to the lower-energy morphology
probed by Fermi-GBM \citep[see][for details]{joint_paper}.

After the detection of GRB~170817A, INTEGRAL continued uninterrupted
observations of the same sky region until 20:44:01 (UTC on August 17). \ch{During this period, no other bursts or steady
  emission from the direction of the optical counterpart were
  detected. We report in detail our flux limits in
  Appendix~\ref{sec:continuation}, while in
  Fig.~\ref{fig:follow_timeline}, we graphically summarize our
  results.}

\begin{figure}
	\centering
	\includegraphics[width=1.\linewidth]{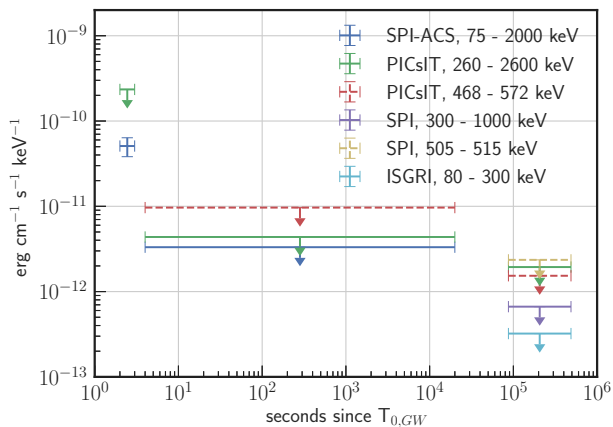}
	\caption{Timeline of the INTEGRAL observations, from the prompt
		detection with SPI-ACS, through the serendipitous
		follow-up and toward the targeted follow-up. Dashed lines
		correspond to the narrowband upper limits. Only selected upper
		limits are shown; for a complete summary of the observations, see
		Table~\ref{tab:fov}, Figure~\ref{fig:spectral_sensitivity}, and
		the text.}
	\label{fig:follow_timeline}
\end{figure}

\section{Targeted INTEGRAL follow-up observation}

\subsection{Search for a Soft Gamma-Ray Afterglow}

INTEGRAL allows us to search for an afterglow emission in a broad
energy range from 3~keV to 8~MeV.  This was covered in detail in
\citet{Savchenko2017a}, where we exploited the serendipitous coverage
of part of the LVT151012 localization region within the field of view
of the INTEGRAL pointed instruments.

To search for a delayed signal, INTEGRAL performed targeted follow-up
observations of the LIGO/Virgo candidate BNS merger G298048
(=GW170817). They started 19.5 hours after the event centered the best
Fermi-GBM localization \citep{GCN21506}. They covered only a
negligible fraction of the LIGO/Virgo localization. Therefore, we avoid
discussing this initial part of the follow-up.

The main part of the follow-up observations was centered on the
candidate optical counterpart, \VAR{optical_counterpart.name}
\citep[RA=13:09:48.089 Dec=-23:22:53.35;][]{GCN21529}. It spanned from
2017 August 18 at 12:45:10 to 2017 August 23 at 03:22:34 (starting about 24
hr after the LIGO/Virgo event), with a maximum on-source time of
326.7 ks. \VAR{optical_counterpart.name} was in the highest
sensitivity part of IBIS and SPI FoV in each of the
dithered single $\sim$40 minute long individual pointings that make
up INTEGRAL observations; \ch{it was in the JEM-X FoV (defined as the
  region where the sensitivity is no less than a factor of 20 from the
  optimal) for
  \VAR{(coverage_pointing.jemx_pointing_fraction_f20*100)|int}\% of
  the time, and
  \VAR{(coverage_pointing.jemx_pointing_fraction_f2*100)|int}\% of the
  time in the region with sensitivity no worse than a factor of 2 from the
  optimal.}

\begin{figure}
  \centering
  \includegraphics[width=1.\linewidth]{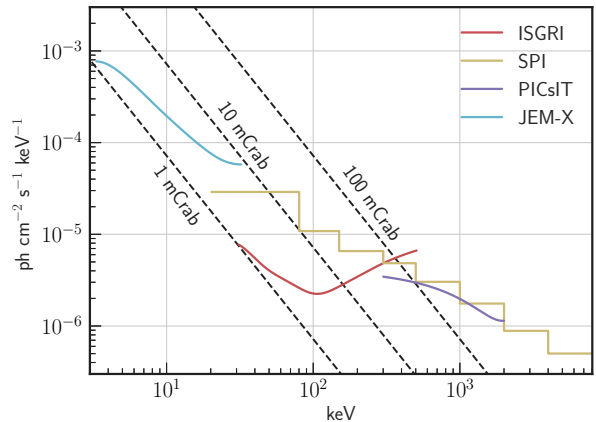}
  \caption{Broadband X-ray to gamma-ray sensitivity reached in the complete INTEGRAL
    targeted follow-up observation, with a total exposure up to 330~ks
    (depending on the instrument and the operational mode).}
\label{fig:spectral_sensitivity}
\end{figure}

We investigated the mosaicked images of the complete observation of
IBIS/ISGRI, SPI, and JEM-X.  We do not detect any X-ray
or gamma-ray counterpart to \VAR{optical_counterpart.name} in any of the instruments. The
3-$\sigma$ broadband upper limits for an average flux of a source at the
position of \VAR{optical_counterpart.name} are presented in
Fig.~\ref{fig:spectral_sensitivity} and in Table~\ref{tab:fov}.

\begin{figure}
  \centering
  \includegraphics[width=1.\linewidth]{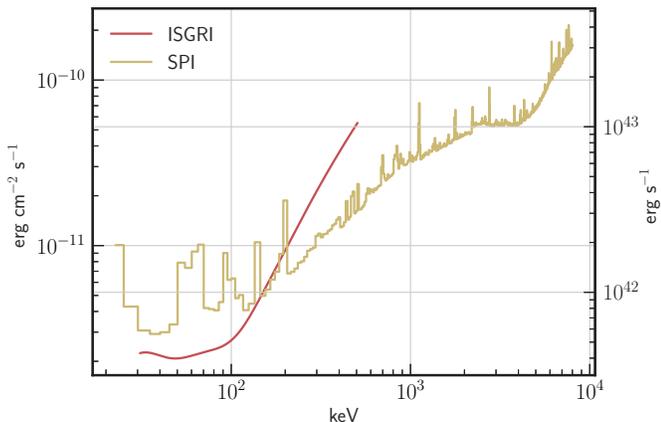}
  \caption{Narrow-line sensitivity in the X-ray/gamma-ray band reached
    in the complete INTEGRAL targeted follow-up observation, with a
    total exposure of \ch{330~ks for each of the instruments. The
      units of the right vertical axis correspond to the luminosity
      assuming a distance to the source of 40\,Mpc.}}
\label{fig:line_sensitivity}
\end{figure}

\begin{table*}[t]
\label{tab:fov}
\centering
\caption{Summary of sensitivities for the different instruments on
  board INTEGRAL to a source at the location of \VAR{optical_counterpart.name}}
\begin{tabular}{ c c c c c c c }

\toprule
  Instrument & Field of View  & Angular resolution & Energy range &   \multicolumn{3}{c}{3~$\sigma$ sensitivity} \\
             &            deg$^2$  &         &                    &          mCrab & erg~cm$^{-2}$~s$^{-1}$ & erg~s$^{-1}$ \\
\\
\midrule

\multirow{2}{*}{JEM-X}& \
\multirow{2}{*}{110} & \
\multirow{2}{*}{3'} & \
     3~--~10~keV & \
     1.2 & \
     1.9$\times$10$^{-11}$ & \
     3.6$\times$10$^{42}$
  \\

      & \
      & \
      & \
 10~--~25~keV & \
 0.64 & \
 7.0$\times$10$^{-12}$ & \
     1.3$\times$10$^{42}$                             
  \\

\midrule

\multirow{3}{*}{IBIS/ISGRI}& \
\multirow{3}{*}{823}& \
\multirow{3}{*}{12'}& \
               20~--~80~keV &\
               2.6 & \
               3.8$\times$10$^{-11}$  &\
     7.3$\times$10$^{42}$                                                             
  \\

 &  \
  & \
  & \
       80~--~300~keV & \
                       6.2 & \
       7.1$\times$10$^{-11}$  &\
     1.4$\times$10$^{43}$                                                             
                                
  \\

    &  \
  & \
  & \
       300~--~500~keV & \
                       290 & \
       1.0$\times$10$^{-9}$  &\
     1.9$\times$10$^{44}$                                                                                             
  \\

\midrule

\multirow{3}{*}{IBIS/PICsIT}& \
\multirow{3}{*}{823}& \
\multirow{3}{*}{24'}& \
               208~--~468~keV &\
               36 & \
                     2.1$\times$10$^{-10}$  &\
     4.0$\times$10$^{43}$                                                             

  \\

              &  \
  & \
  & \
       468~--~572~keV & \
                       128 & \
       1.6$\times$10$^{-10}$  &\
                           3.1$\times$10$^{43}$                                                             
          
  \\

             &  \
  & \
  & \
       572~--~1196~keV & \
                       216 & \
       8.7$\times$10$^{-10}$  &\
                           1.7$\times$10$^{44}$                                                             
          
  \\

               &  \
  & \
  & \
       1196~--~2600~keV & \
                       973 & \
       3.3$\times$10$^{-9}$  &\
     6.4$\times$10$^{44}$                                                             
                                
  \\

\midrule

  \multirow{6}{*}{SPI}& \
\multirow{6}{*}{794}& \
\multirow{6}{*}{2.5$^\circ$}& \
               20~--~80~keV &\
               3.8 & \
               5.6$\times$10$^{-11}$  &\
     1.1$\times$10$^{43}$                                                             
                                
  \\

& \
& \
& \
               80~--~150~keV &\
               16.4 & \
               9.8$\times$10$^{-11}$  &\
           1.9$\times$10$^{43}$                                                             
                          
  \\

  & \
& \
& \
               150~--~300~keV &\
               43 & \
               2.4$\times$10$^{-10}$  &\
               4.6$\times$10$^{43}$                                                             
                                
  \\

   & \
& \
& \
               300~--~500~keV &\
               135 & \
               4.7$\times$10$^{-10}$  &\
                   9.0$\times$10$^{43}$                                                             
                  
  \\

   & \
& \
& \
               500~--~1000~keV &\
               308 & \
               1.2$\times$10$^{-9}$  &\
                    2.3$\times$10$^{44}$                                                             
                 
  \\

   & \
& \
& \
               1000~--~2000~keV &\
               866 & \
                     2.8$\times$10$^{-9}$  &\
     5.4$\times$10$^{44}$

  \\

   & \
& \
& \
               2000~--~4000~keV &\
               \VAR{table.spi[2000,4000].mcrab|int} & \
               \VAR{table.spi[2000,4000].flux_ecs|latex_exp} & \
               \VAR{table.spi[2000,4000].L|latex_exp} 
                                             
  \\

   & \
& \
& \
               4000~--~8000~keV &\
               \VAR{table.spi[4000,8000].mcrab|int} & \
               \VAR{table.spi[4000,8000].flux_ecs|latex_exp} & \
               \VAR{table.spi[4000,8000].L|latex_exp} 
                                             
  \\

\bottomrule
\vspace{0.3cm}
\end{tabular}

\begin{tablenotes}
\item The upper limits from the INTEGRAL follow-up observation
  directed towards \VAR{optical_counterpart.name}, assuming a
  power-law-shaped spectral energy distribution with a photon index of
  $-2$. The energy ranges are chosen to highlight the advantage of
  INTEGRAL instruments over other hard X-ray observatories. The limit
  of the FoVs has been set to the point where a worsening of the
  instrument sensitivity by a factor of 20 compared to the on-axis
  value is reached.
    \end{tablenotes}
\end{table*}

We have also searched for isolated line-like features in IBIS/ISGRI
and SPI data: our preliminary analysis did not identify any such
features. In-depth studies will be reported elsewhere.  The
narrow-line sensitivity reached in the complete follow-up observation is
presented in Figure~\ref{fig:line_sensitivity}.

\begin{figure}
  \centering
  \includegraphics[width=1.\linewidth]{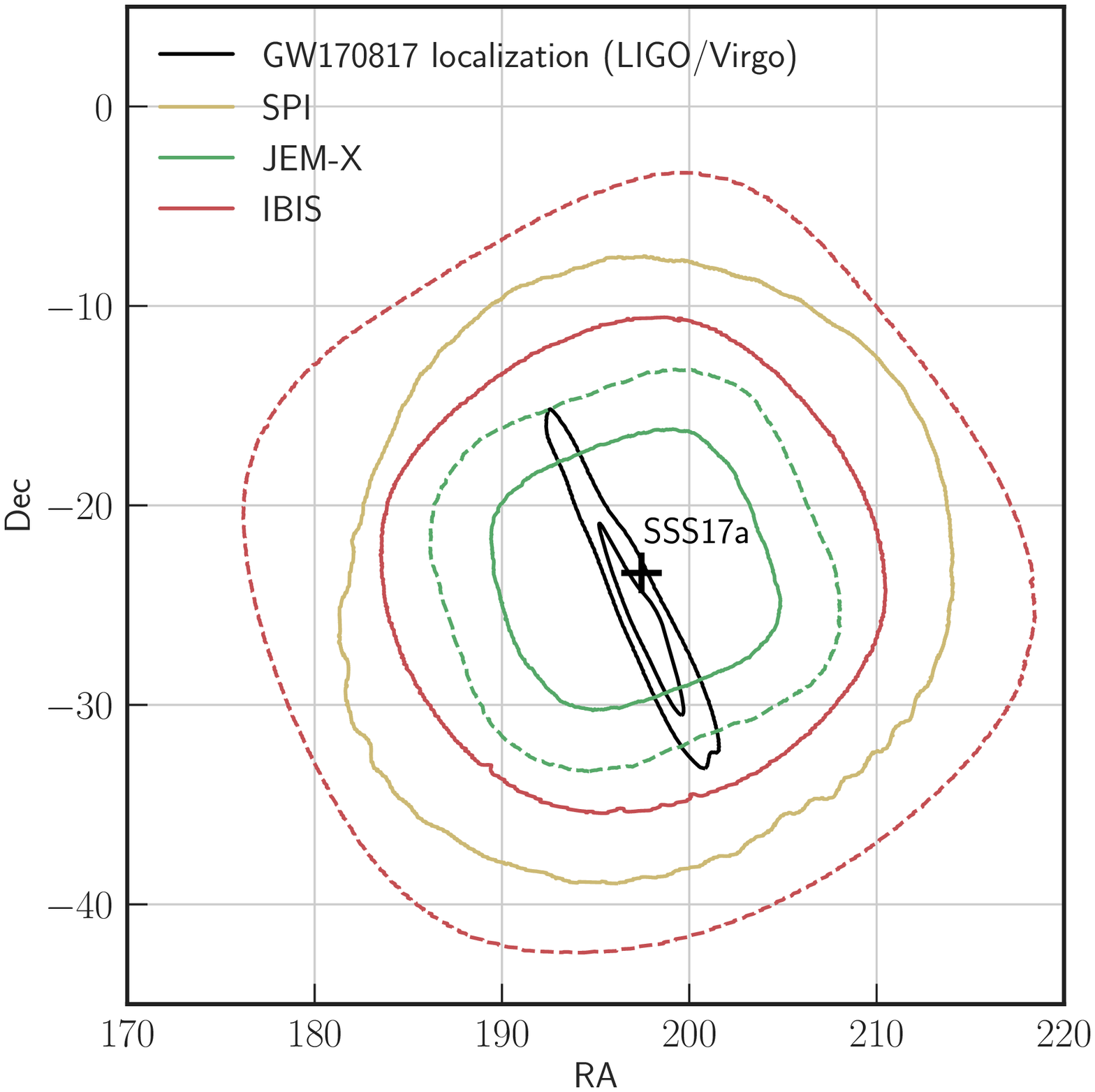}
  \caption{Sensitivity levels (50\% - solid line, and 10\% - dashed
    line, of the optimal sensitivity, achieved for \VAR{optical_counterpart.name}) of the
    complete IBIS, JEM-X, and SPI mosaics of the targeted INTEGRAL follow-up
    observation, compared to the most accurate LALInferrence
    LIGO/Virgo localization of GW170817 (50\% and 90\% confidence
    containment \citep[black solid lines][]{GCN21527}) and the \VAR{optical_counterpart.name} location
    \citep{GCN21529}.}
\label{fig:follow_coverage}
\end{figure}

IBIS, SPI, and JEM-X observed more than 97\% of the LIGO-Virgo
localization in the combined observation mosaic.  We searched the
IBIS/ISGRI, SPI, and JEM-X data for any new point source in the whole
90\% LIGO/Virgo localization region, and did not find any. The
sensitivity depends on the location, with the best value close to that
computed for \VAR{optical_counterpart.name}. Contours containing
regions observed with sensitivity of at least 50\% and 10\% of the
optimal are presented in Figure~\ref{fig:follow_coverage}).

\subsection{Search for optical emission with the OMC}

The OMC observed the galaxy NGC 4993 including the transient \VAR{optical_counterpart.name}
from 2017 August 18 at 17:27:59 until 2017 August 22 at 22:56:48 UTC. It was
in the OMC FoV for only 29 INTEGRAL pointings (total of 35.7\,ks). Its limited angular
resolution and pixel size did not allow us to distinguish between the
host galaxy contribution and the transient. The data were analyzed by
using the largest OMC photometric aperture (5x5 pixel, 90~arcsec
diameter), to ensure all of the emission from the host galaxy as well as
the transient are included in the aperture. We measured a V magnitude
of 12.67$\pm$0.03 (1-$\sigma$ level) for the total emission. No
variability was detected in the OMC data at the reported 1-sigma
level.

\subsection{Search for delayed bursting activity}\label{sec:delayedbursting}

The continuous observation of the \VAR{optical_counterpart.name}
location performed by INTEGRAL (from 2017 August 18 at 12:45:10 to
2017 August 23 at 03:22:34 UT with a coverage fraction of 80\%) allows us
to also search for any short (magnetar-like) or long bursts from this
source. We used IBIS/ISGRI light curves in two energy ranges:
20--80\,keV and 80--300\,keV, on 100\,ms, 1\,s, 10\,s, and 100\,s time
scales. We did not find any deviations from the background, and set a
3~$\sigma$ upper limit on any possible 1-s long burst flux of 1.0 Crab
(1.4$\times10^{-8}$\,erg\,cm$^{-2}$\,s$^{-1}$) in the 20--80\,keV, and
6.8 Crab (7.8$\times10^{-8}$\,erg\,cm$^{-2}$\,s$^{-1}$) in the
80--300\,keV energy range. The 3~$\sigma$ upper limit on a 100-ms time
scale results in 3.0 Crab
(4.5$\times10^{-8}$\,erg\,cm$^{-2}$\,s$^{-1}$) in the 20--80\,keV, and
21 Crab (2.4$\times10^{-7}$\,erg\,cm$^{-2}$\,s$^{-1}$) in the
80--300\,keV energy range. Assuming a distance of 40\,Mpc \ch{(see
  online data of \citealt{Crook2007})}, these limits can be
interpreted as constraints on the burst luminosity on 1~s (100~ms)
time scale of 2.6$\times$10$^{45}$\,erg\,s$^{-1}$
(8.5$\times$10$^{45}$\,erg\,s$^{-1}$) in the 20--80\,keV energy
range. In the 80--300\,keV energy range the luminosity is constrained
to be less than 1.5$\times$10$^{46}$\,erg\,s$^{-1}$
(4.6$\times$10$^{46}$\,erg\,s$^{-1}$). Bursts exceeding such
luminosities were previously observed from magnetars. SGR~1806$-$20,
for example, produced a giant flare with a total energy of
2$\times$10$^{46}$~erg \citep{Hurley2005}, while individual flares
from, e.g., 1E~1547.0$-$5408 exceeded the energy of 10$^{46}$~erg
\citep{Mereghetti2009,Savchenko2010}.

\section{Discussion}\label{sec:discussion}

The detection of GRB~170817A by INTEGRAL and Fermi in unambiguous
coincidence with GW170817 is the first definitive proof that at least
some sGRBs can be associated with BNS merger events. \ch{The duration of the GRB as measured by INTEGRAL above $\sim$100\,keV, 100~ms, firmly assigns
the GRB~170817A to the short GRB class.} \citet{joint_paper} extensively
discuss the implications of the joint GW and gamma-ray observation for the
luminosity function and structure of the sGRB population.

Future observations of similar events will be decisive in constraining
the properties of the BNS merger counterparts.  INTEGRAL exhibits an
exceptionally unocculted ($>$99.9\%) sky view at every moment when it
is observing, i.e., for about 85\% of the total mission lifetime. With
its high sensitivity above $\sim$100~keV, INTEGRAL/SPI-ACS has
demonstrated its capability of detecting also fairly weak and soft
transients like GRB~170817A.  In the future, INTEGRAL will be able to
systematically detect counterparts to the GW events or to put tight
upper limits on their presence. The advantage of its exceptionally
high effective area above $\sim$100~keV will be even more important
for the events with harder spectra, which is what expected for
typical \acp{sGRB}.

\ch{The possibility of forming a (short-lived) magnetar in a BNS
  merger has been extensively discussed in the past
  \citep[e.g.][]{Metzger2014,Giacomazzo2015,Price2006,Fernandez2016}. It
  has been suggested that a newborn magnetar is responsible for part
  of the afterglow emission \citep[e.g.][]{Rowlinson2013}.  In
  principle, it is not clear if a newborn magnetar, formed in the
  merger, is more likely to produce bursts during the first days after
  the merger (which are covered by the continuous INTEGRAL
  observation). This could be reasonably expected because at early
  times, the magnetic energy is maximal and it rapidly
  dissipates. Intense X-ray flares could occur associated to frequent
  reorganization in the magnetic field structure, as well as in
  connection to a delayed accretion.}  We have ruled out, however, the
existence of strong magnetar-like bursts in our targeted follow-up
observation.

Interestingly, the amount of energy released in GRB~170817A
\citep{joint_paper} is similar to that found during the giant flares
of magnetars, such as SGR~1806$-$20 \citep{Hurley2005}. \ch{Magnetar
  flares are associated with long-lived objects while the GRB~170817A
  was a one-time event of a BNS merger. Nevertheless, the similarity
  in the most basic observational properties is intriguing and it may
  point towards a similarity of the physical processes involved.}

\ch{At late times after the initial gamma-ray burst}, the luminosity of kilonovae is largely fueled by radioactive decays of
r-process elements released in the coalescence \citep[see, e.g.,][for a review on kilonova mechanism]{Metzger2017}. Under favorable
conditions, a forest of nuclear gamma-ray lines produced in these
decays may be detectable by a suitable gamma-ray spectrometer such as
INTEGRAL/SPI \citep{Hotokezaka2016}.  If the lines are broad or appear
at low energies ($<$100 keV), IBIS could also detect them, \ch{with a similar significance}. We did not
find any such emission feature 
with INTEGRAL/SPI or IBIS, and we set an upper
limit as displayed in Fig.~\ref{fig:line_sensitivity} \ch{for narrow lines. However, for
sufficiently broad lines, the emission pattern can resemble a
nearly continuous spectrum \citep[as discussed in][for high-velocity
ejecta]{Hotokezaka2016}. Thus, 
the continuum emission upper limit can be applied (see
Fig.~\ref{fig:spectral_sensitivity} and Table~\ref{tab:fov}).}

\ch{To the best of our knowledge, the most favorable predictions for a
combined decay line flux 1~day after the merger and at the
\VAR{optical_counterpart.name} distance are of the order of
\ecse{3.6}{-12} \citep[assuming high ejecta mass of 0.1\,M$_\odot$
and a velocity of 0.3\,c][]{Hotokezaka2016} in the 300~keV~--~1~MeV band;
this is considerably below our best upper limit in the same energy range, i.e.,
\ecse{1.7}{-9}.}

The detectability of gamma-ray emission resulting from the e$^{+/-}$
annihilation strongly depends on the final photon spectrum, which is
in turn determined by the conditions in which annihilation occurs. The
final spectrum could in principle include a narrow or a broad, blueshifted
or redshifted line-like feature near 511~keV, or may be dominated by
a very extended excess in the soft gamma-ray energy range
\citep[e.g.][]{Maciolek1995,Svensson1987}.  To give a general idea
about the sensitivity of INTEGRAL, we consider the IBIS/PICsIT upper
limits in the energy range 468--572~keV, i.e., around the 511~keV
annihilation line, during the targeted follow-up observation. This
limit corresponds to 3.1$\times$10$^{43}$erg~s$^{-1}$ (see
Table~\ref{tab:fov}). This luminosity roughly constraints the total
rate of annihilation to less than
1.7$\times$10$^{-13}$M$_\odot$~s$^{-1}$. A particularly stringent
upper limit can be set by SPI on the flux of a narrow annihilation
line between 505 and 515\,keV, which is less than
4.5$\times$10$^{42}$\,erg\,s$^{-1}$.
 
\section{Conclusions}

We reported the independent INTEGRAL detection of a short gamma-ray
burst (GRB~170817A), in coincidence with that found by Fermi-GBM
\ch{(the association significance between INTEGRAL and Fermi-GBM is
  \VAR{search.assoc.gbm.sig|round(1)} sigma)}, which is for the first
time unambiguously associated to the gravitational wave event GW170817 observed by LIGO/Virgo and consistent with a binary neutron star
merger. \ch{The significance of association between the independent INTEGRAL GRB detection and GW170817 is
\VAR{search.assoc.ligo.sig|round(1)}~sigma}. This is a turning point
for multi-messenger astrophysics.

This observation is compatible with the expectation that a large
fraction (if not all) BNS mergers might be accompanied by a prompt
gamma-ray flash \citep{joint_paper}, detectable by INTEGRAL/SPI-ACS
and other facilities.  INTEGRAL independently detects more than 20
confirmed sGRBs per year \citep{Savchenko2012}) in a broad range of
fluences. With the growing sensitivity of the LIGO and Virgo
observatories, being joined in the future by other observatories, we
expect to detect more and more short GRBs associated with BNS mergers.

Additionally, we have exploited the unique uninterrupted serendipitous
INTEGRAL observations available immediately after GRB~170817A/GW170817 (lasting
about 20~ks), as well as dedicated targeted follow-up observations
carried out by INTEGRAL, starting as soon as 19.5 hours after the GRB/GW
(lasting in total 5.1 days). No hard X-ray or gamma-ray
signal above the background was found. By taking advantage of the full
sensitivity and wide FoV of the combination of the IBIS, SPI, and
JEM-X instruments, we provide a stringent upper limit over a broad
energy range, from 3\,keV up to 8\,MeV. The INTEGRAL upper limits above
80\,keV are tighter than those set by any other instrument and
constrain the isotropic-equivalent luminosity of the soft gamma-ray
afterglow to less than 1.4$\times$10$^{43}$~erg~s$^{-1}$
(80--300\,keV), assuming a distance of 40\,Mpc to the source   Our data exclude the possibility that a short- or a
long-lasting bright hard X-ray and/or soft gamma-ray phase of activity followed
GRB~170817A/GW170817.

With these results, we show that 
INTEGRAL continues to play a key role in the rapidly emerging
multi-messenger field by constraining both the prompt and
delayed gamma-ray emission associated with compact object mergers.

\section*{Acknowledgments}

This work is based on observations with INTEGRAL, an ESA project with
instruments and science data center funded by ESA member states
(especially the PI countries: Denmark, France, Germany, Italy,
Switzerland, Spain), and with the participation of Russia and the
USA. The INTEGRAL SPI project has been completed under the
responsibility and leadership of CNES. The SPI-ACS detector system has
been provided by MPE Garching/Germany. The SPI team is grateful to
ASI, CEA, CNES, DLR, ESA, INTA, NASA and OSTC for their support. The
Italian INTEGRAL team acknowledges the support of ASI/INAF agreement
n.\ 2013-025-R.1. RD and AvK acknowledge the German INTEGRAL support
through DLR grant 50 OG 1101. AL and RS acknowledge the support from
the Russian Science Foundation (grant 14-22-00271). AD is funded by
Spanish MINECO/FEDER grant ESP2015-65712-C5-1-R. Some of the results
in this paper have been derived using the \software{HEALPix}
\citep{healpix} package. We are grateful the Fran\c cois Arago Centre
at APC for providing computing resources, and VirtualData from LABEX
P2IO for enabling access to the StratusLab academic cloud.  We
acknowledge the continuous support by the INTEGRAL Users Group and the
exceptionally efficient support by the teams at ESAC and ESOC for the
scheduling of the targeted follow-up observations. We are grateful to
the LVC and Fermi-GBM teams for their suggestions on earlier versions
of this Letter. Finally, we thank the anonymous referee for
constructive suggestions.

\appendix

\section{Association significance}
\label{sec:association}

It is well known that cosmic-ray (CR) interactions can
cause short spikes in the ACS light curve \citep{Savchenko2012}. The
false-alarm rate (FAR) in SPI-ACS for short and weak events, like
GRB~170817A, is largely determined by the ability to discriminate
between CR-induced and astrophysical events.  Following \cite{Savchenko2012},
we exploit the universality of the CR-induced spike temporal profile
in SPI-ACS and compute the significance of an event to adhere to the
NULL hypothesis of having a spike-like profile; we then multiply this
number by the S/N of the event to obtain a numeric ranking.

We used data from the same INTEGRAL spacecraft revolution (2017 August 15
at 15:27:42 -- 2017 August 17 at 18:26:32 UTC, 51~hours in total) to
compute the FAR for the events with a rank not smaller than that of
GRB~170817A.  This results in 2.2$\times$10$^{-5}$~Hz and can be used
to compute a post-trial false alarm probability (FAP) for an excess at
T$_{0,ACS}$ to be associated with the GW trigger as P =
2$\times$~2.2$\times$10$^{-5}$~Hz~$\times$~2~s~$\times$(1
+$\log\left(30\mathrm{\,s / 0.1\, s}\right)$)$\sim$ 0.07\% (3.2~sigma). Here,
we use the 2~s time difference with the association target (GW170817),
the 30~s half time window of the search and the 0.1~s for the phase
step in the minimal search time scale. The factor of two is due to
counting both before and after the trigger.

The evidence for association with the Fermi-GBM GRB detection can be
derived by assuming an association time difference of 50~ms. This is a
conservative value taking into account the time bin of SPI-ACS and a
marginally allowed offset between the temporal profiles of two events,
with the light travel time correction assuming the location of
\VAR{optical_counterpart.name}. This results in an association
significance of \VAR{search.assoc.gbm.sig|round(1)}~sigma, providing strong evidence that the event
detected by INTEGRAL/SPI-ACS is associated to the GRB~170817A, detected
by the Fermi-GBM on-board algorithms. It is interesting to
derive also an association significance using only the Fermi-GBM and
INTEGRAL/SPI-ACS results, i.e., using the GBM location of GRB~170817A
for the light travel time correction. The extent of the localization
region derived from the GBM data alone accounts for an additional time
uncertainty of 100~ms, resulting in an association significance of
3.9~sigma.

\section{Search for prompt emission with IBIS and SPI}
\label{sec:prompt_multiinstrument}

We have inspected data of the IBIS/ISGRI, IBIS/PICsIT, and SPI
main detectors, and did not find any excess at the T$_{0,ACS}$. Even
though none of these instruments had substantial sensitivity in the
direction of \VAR{optical_counterpart.name} (see
\citealt{Savchenko2017a} for the all-sky angular dependency of
sensitivity of different instruments), this inspection is particularly
interesting, because particle background variations tend to appear in
multiple instruments at once. The non-observation of any excess in
instruments where we do not expect an astrophysical signal disfavors
the hypothesis that local particle background excess contributed to
the event, and further supports that our detection is associated with
\VAR{optical_counterpart.name} and the GW event. Note that we did not
use the X-ray information from the JEM-X instrument, since in its
energy range ($<$30~keV) it was heavily shielded from any emission in
the direction of \VAR{optical_counterpart.name}.

\section{Search for an early soft gamma-ray afterglow}
\label{sec:continuation}
After the detection of GRB~170817A, 
INTEGRAL continued to point toward
the same sky region (i.e., essentially with no change to the response
for the position of \VAR{optical_counterpart.name}) in stable background conditions, until the
instruments were switched off for the perigee passage at 20:44:01 (UTC
on August 17). The total duration of this observation was about 20~ks.
Exploiting the all-sky sensitivity of SPI-ACS and IBIS/PICsIT, we
derive upper limits on any early gamma-ray afterglow for this period
of time.

Using the background rate from the earlier part of the INTEGRAL spacecraft 
revolution 1851 (the revolution that contains the GRB~170817A) and
the data of neighboring revolutions, we estimate a 3~sigma upper limit
on the average flux of any new source in the single broad energy range
75--2000\,keV accessible to SPI-ACS of 280~mCrab or
1.2$\times$10$^{45}$~erg~s$^{-1}$, assuming a Crab-like spectrum.

While the response of IBIS/PICsIT to a GRB-like spectrum in the
direction of \VAR{optical_counterpart.name} was less than optimal, the energy resolution in
the PICsIT spectral-timing mode allows us to set a constraining upper
limit on emission limited to a narrow range of gamma-ray energies.
The 3-sigma upper limits on a source flux average over the studied
time interval exploiting PICsIT have been determined in three energy
bands.  Two broad ones and a narrow one centered on 511\,keV to search
for broad line emission.  The upper limits from these serendipitous
observations are: 2.8$\times10^{45}$\,erg\,s$^{-1}$
(260--468\,keV), 3.1$\times10^{44}$\,erg\,s$^{-1}$
(468--572\,keV), and 3.3$\times10^{45}$\,erg\,s$^{-1}$
(572--2600\,keV).

Using the SPI-ACS data in the same time interval, we also searched for
isolated short bursts on timescales from 100~ms to 10~s. We do not
find any evidence for further bursting activity and set an upper limit
on any excess at the level of $\sim$5~$\times$10$^{-7}$~erg~cm$^{-2}$ on 1~s
time scale for the 75\,keV--2\,MeV energy range.

\bibliographystyle{aa}
\bibliography{references}

\end{document}